\documentclass[preprint,authoryear,12pt]{elsarticle}

\usepackage{epsfig}

\usepackage{amssymb}

\usepackage[ps2pdf,%
a4paper=true,%
breaklinks=true,%
colorlinks=true,%
pdfauthor={De la Luz et al.},%
pdftitle={Solar Limb}%
]{hyperref}

\usepackage{natbib}
\newcommand{\solphys}{{\it Solar Phys.}}

\journal{Advances in Space Research}

\begin{document}

\begin{frontmatter}



\title{Solar Limb Theoretical Tomography at Millimeter, Sub-millimeter, and Infrared Wavelengths}


\author{Victor De la Luz}
\cortext[cor]{Corresponding author}

\ead{vdelaluz@geofisica.unam.mx}

  \author{J. A. Gonzalez-Esparza, P. Corona-Romero, and J. Mejia-Ambriz}

\address{SCiESMEX, Instituto de Geofisica, Unidad Michoacan, Universidad Nacional Autonoma de Mexico, Morelia, Michoacan, Mexico. CP 58190.}


\begin{abstract}
Semi-empirical models of the solar Chromosphere show in their emission spectrum, tomography property at millimeter, sub-millimeter, and infrared wavelengths for the center of the solar disk. In this work, we studied this property in the solar limb using our numerical code PakalMPI, focusing in the region where the solar atmosphere becomes optically thick. Individual contribution of Bremsstrahlung and H- opacities was take into account in the radiative transfer process. We found that the tomography property remains in all the spectrum region under study at limb altitudes. For frequencies between 2 GHz and 5 THz the contribution of Bremsstrahlung is the dominant process above the solar limb. 
\end{abstract}


\end{frontmatter}

\parindent=0.5 cm

\section{Introduction}
The limb radio emission of the Quiet Sun is characterized by a strong brightening at low frequencies \citep{1947RSPSA.190..357M}. Early studies to characterize the solar limb brightening found that this strong emission could extend until sub-millimeter wavelengths in the electromagnetic spectrum \citep{1995ApJ...453..511L}. Theoretical modeling of the emission in the solar limb involves semi empirical models of the Chromosphere by focusing in the center of the solar disk \citep{1981ApJS...45..635V,2008ApJS..175..229A,2012SoPh..277...31H}. However, observations with high spatial resolution of the solar limb shows significant differences at low frequencies with respect to semi empirical models \citep{1994IAUS..154..139C}.

If we analyze the Chromospheric emission in terms of the optical depth, we found that the atmosphere is observable at the frequency $\nu$ only for those altitudes higher than the region where $\tau_{\nu} <= 1$. In this sense, using direct observations, $\tau_{\nu} \approx 1$ is an optical border to analyze the physical process in the radiative transfer \citep{2014SoPh..289.2879D}. The height where $\tau_{\nu}\approx 1$ depends on the frequency; in general, as higher the frequency 
as closer to the photosphere is the source of the emission. This is the tomography property of the Chromosphere at millimeter - infrared wavelength region \citep{2011SoPh..273..309S}. At different altitudes, it corresponds to different emissions and absorption processes, the observational $\nu$ range is related to the physical mechanism that originates the radiation.

In the center of the solar disk, \cite{2011ApJ...737....1D} found that for altitudes higher than $500$ km above the photosphere the Bremsstrahlung is the predominant process for the millimeter - infrared range. However, for altitudes lower than $500$ km, H- becomes the dominant process in the radiative transfer solution. 

In this work, we analyzed the optical depth in the solar limb between 2 GHz and 10 THz to characterize the tomography property observed in the center of the solar disk but focused at limb altitudes. We also studied the physical processes at Chromospheric altitudes at the limb using the Chromospheric C07 model \citep{2008ApJS..175..229A}.

The order of the paper is as follows: in section 2 we describe the geometry, in section 3 we introduce the Chromospheric model and the radiative transfer solution. Section 4 address the spectrum computations. Section 5 presents the results and Section 6 the Conclusions.

\section{The Geometry}
\begin{figure}
\begin{center}
\includegraphics*[width=1.0\textwidth]{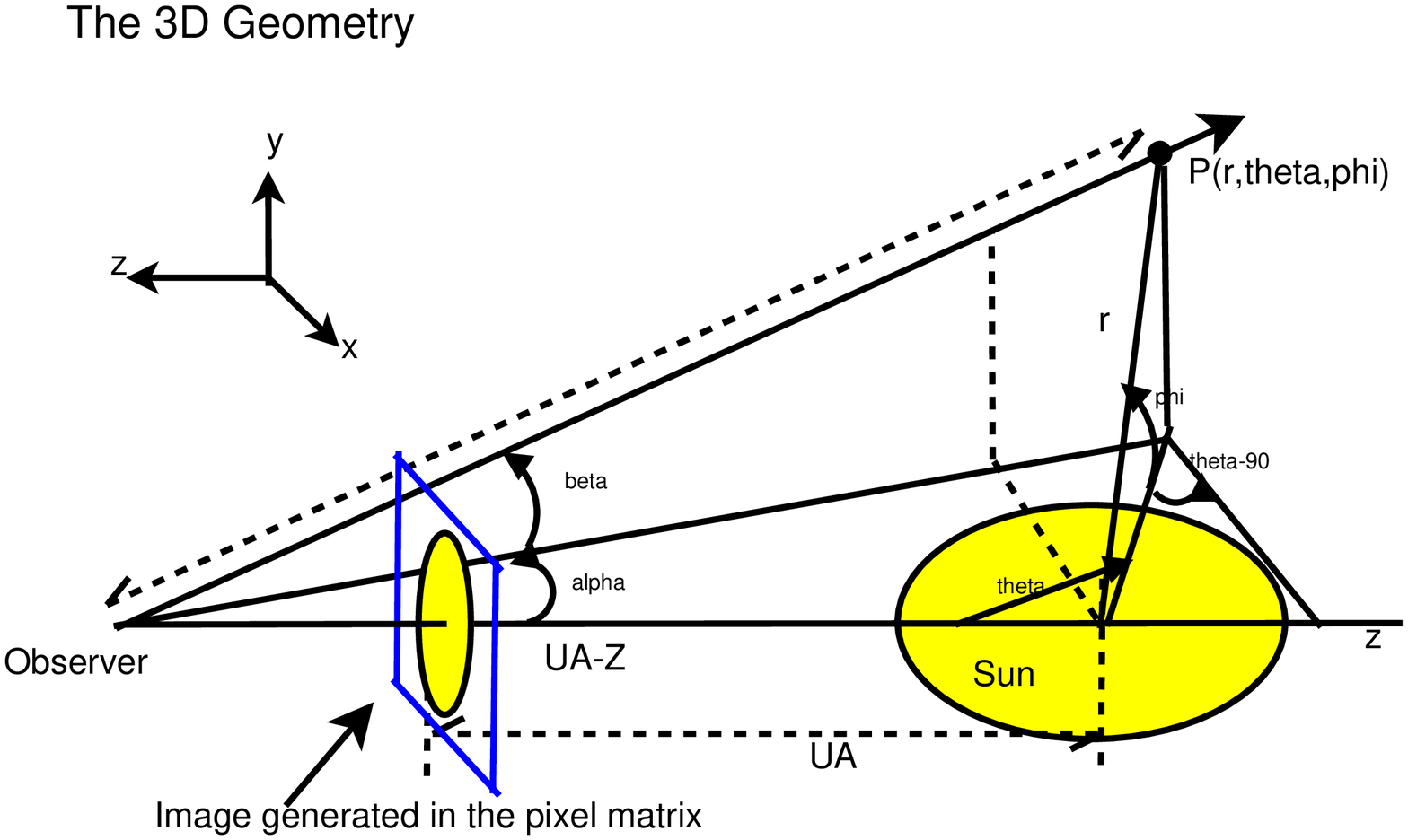}
\end{center}
\caption{3D geometry using vanishing point configuration. The observer is located at 1UA from the Sun. }\label{colorgeometriaEN.eps}

\end{figure}

\begin{figure}
\begin{center}
\includegraphics*[width=1.0\textwidth]{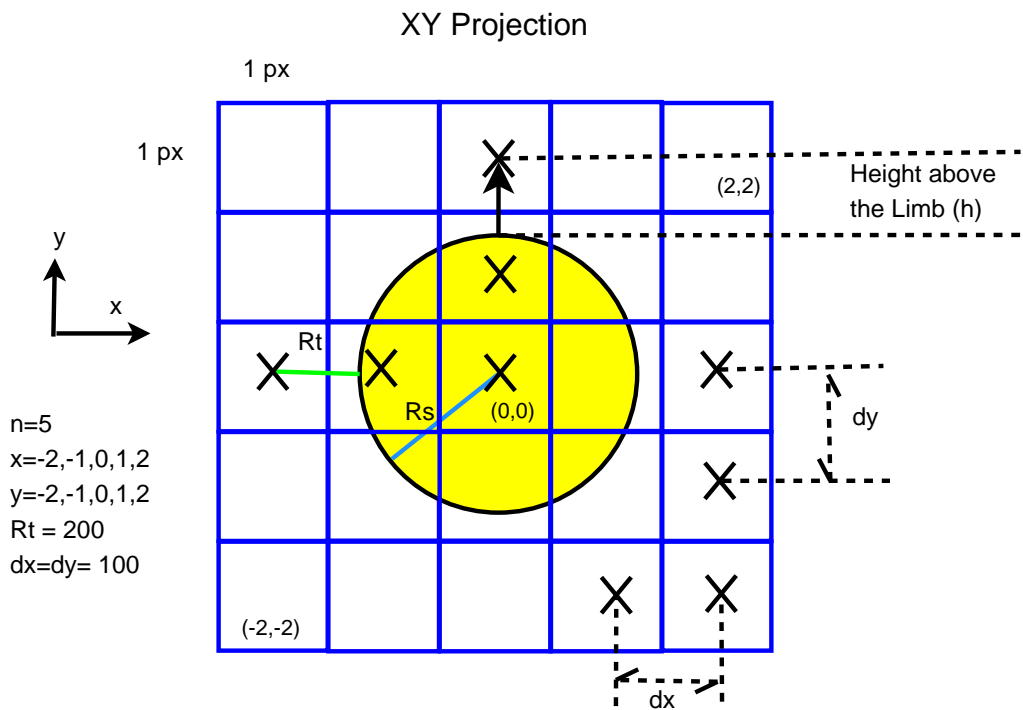}
\end{center}
\caption{Projection of the geometry in the XY plane. Note that the height above the limb starts at 1 $R_{sun}$ in the y axis. The physical length of each pixel is defined by the resolution of the image (n) and the variable Rt. }\label{colorgeometriaEN-2.eps}
\end{figure}

In Figure \ref{colorgeometriaEN.eps}, we introduce the 3D geometry used in this work (for more information see \cite{2010ApJS..188..437D}). The geometry is characterized by the resolution of the image and the zoom parameter (Rt). The geometry is based in the principle of the vanishing point. The observer is located in the Earth at 1 UA from the Sun. The origin of our geometry is the center of the Sun and the z-axis is defined from the Sun to the observer. Using Rt and the resolution in pixels of the image (n) we can compute the angles alpha and beta related for each pixel in the 2D projection on the XY plane (Figure \ref{colorgeometriaEN-2.eps}). For each alpha and beta angles, we computed the set of points $P$ in one line of sight. Each point in the line of sight is related to the distance $r$ (distance between the center of the Sun and $P$). Finally, with $r$ we recovered the temperature and the density from the radial models.

In this work, we defined the ``heigh above the limb (h)'' as the distance over the Y-axis between the border of the Sun in the XY projection and this point.

\section{The Model}
Taking into account the 3D geometry at the solar limb, we used the C07 Chromospheric semi empirical model \citep{2008ApJS..175..229A} as atmospheric input model of temperature and density. The code PakalMPI solves the radiative transfer equation \citep[RTE][]{2010ApJS..188..437D}. The C07 model is fully tested in an wide range of frequencies in the continuum and in emission lines.
The PakalMPI code solves the radiative transfer equation iteratively using Message Passing Interface (MPI) to compute the different ray paths in parallel \citep{2010ApJS..188..437D}. The code computes the electronic density, Hydrogen, and H- in Non Local Thermodynamic Equilibrium together with He and 18 atoms in LTE using the Saha equation.
In this work, we used two opacity functions: Classical Bremsstrahlung and H- \citep{2011ApJ...737....1D}. 

\section{Computations}
We computed 2800 spectrums from 2 GHz to 10 THz. The set of spectrums was computed over the Y-axis from  $696$ Mm (1 $R_{sun}$) to $1413$ Mm ($717$ Mm above the solar limb).

For each altitude we solved the radiative transfer equation over the z-axis to produce the final Brightness Temperature. The detailed information of the radiative transfer equation over the z-axis was stored in our database. With this information we computed for each altitude above the limb ($y$) and frequency ($\nu$) the point in the z-axis where $\tau_{\nu,y}(z) = 1$. This point is referred to as $z_0(\nu,y)$, i.e. the distance from the Sun center to the observer where the atmosphere becomes optically thick for an specific frequency $\nu$ and altitude above the limb $y$. We also saved $\kappa_{\nu,y}(z_0)$ for comparing Bremsstrahlung and H- separately to further analysis.

\section{Results}
\begin{figure}
\begin{center}
\includegraphics*[width=1.0\textwidth]{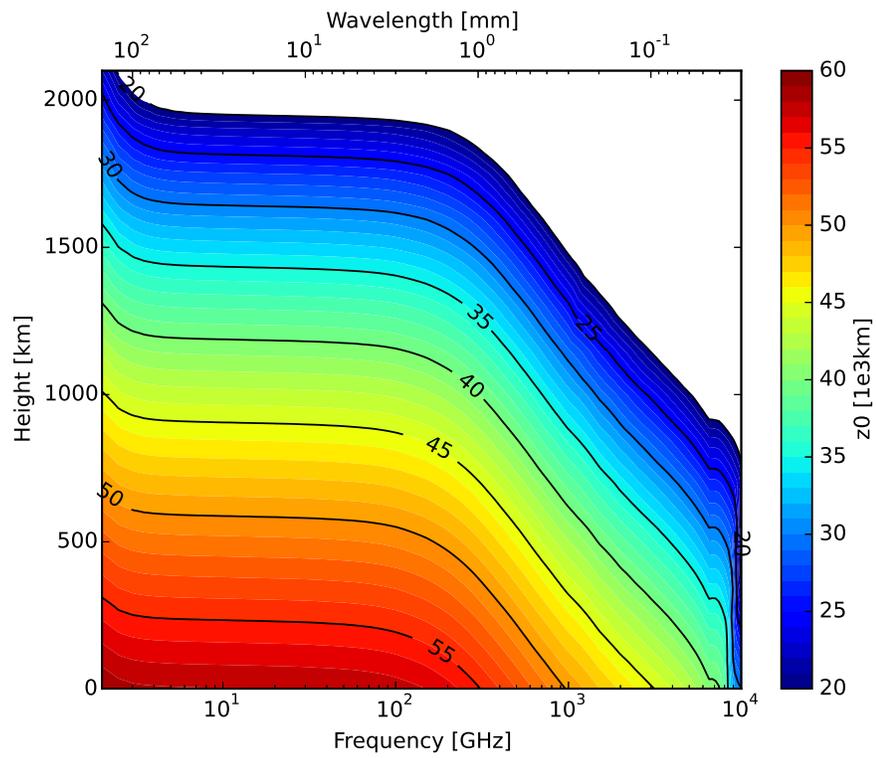}
\end{center}
\caption{Height above the limb versus frequency, in colors $z_0$ (distance in z-axis where $\tau(z_0) = 1$). This plot only shows $z_0 > 20 Mm$. }\label{plot-h-vs-nu-vs-tau1-20-100km.eps}
\end{figure}

Figure \ref{plot-h-vs-nu-vs-tau1-20-100km.eps} shows the computed $z_0(\nu,y)$ (in colors) between 2 GHz and 10 THz (X-axis in the plot) versus altitudes above the limb (Y-axis) between $0$ and $2100$ km. This figure shows the optically thick borders with respect frequency and altitude above the limb. At high frequencies the values of $z_0(\nu,y)$ converges to $0$. This plot provide us the distances over the z-axis where the atmosphere becomes optically thick for both: frequency and altitudes above the limb. For example, for the case of $500$ km above the limb, we observed that $z_0 \approx 52$ km remains constant for frequencies between 2 and 200 GHz. At frequencies higher than 200 GHz $z_0$ drops to lower values. We observed similar behaviour for all frequencies and altitudes under study.  The Figure \ref{plot-h-vs-nu-vs-tau1-20-100km.eps} shows the tomographic property of the solar limb at millimeter - infrared range.

\begin{figure}
\begin{center}
\includegraphics*[width=1.0\textwidth]{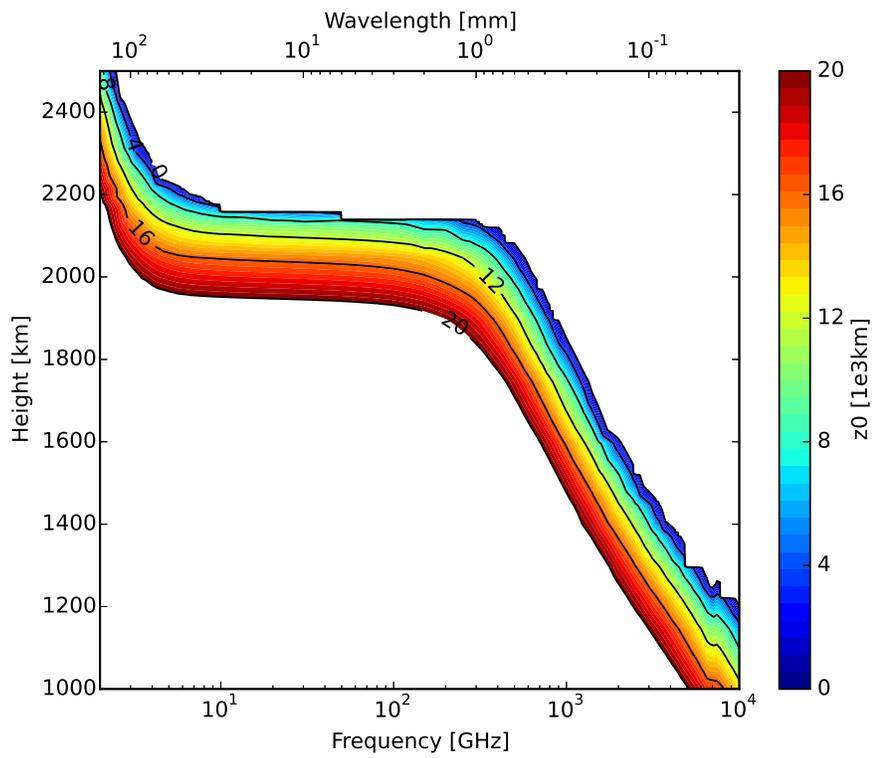}
\end{center}
\caption{Height above the limb versus frequency, in colors $201$ Mm $> z_0 > 0$ km. For $z_0 < 0$ km values of $\tau(z)$ do not reach 1. For this altitudes the atmosphere is optically thin.}\label{plot-h-vs-nu-vs-tau1--20-20km.eps}
\end{figure}

The same case for the Figure \ref{plot-h-vs-nu-vs-tau1--20-20km.eps}, in this figure we plotted $z_0(\nu,y)$ between $0$ km and $20$ km (in z-axis). For $z_0(\nu,y) < 0$ the value of $\tau_{\nu}(z)$ never reaches 1, i.e. the atmosphere is optically thin for all the line of sight. This figure shows the theoretical observational limit in the frequency range under study.

\begin{figure}
\begin{center}
\includegraphics*[width=1.0\textwidth]{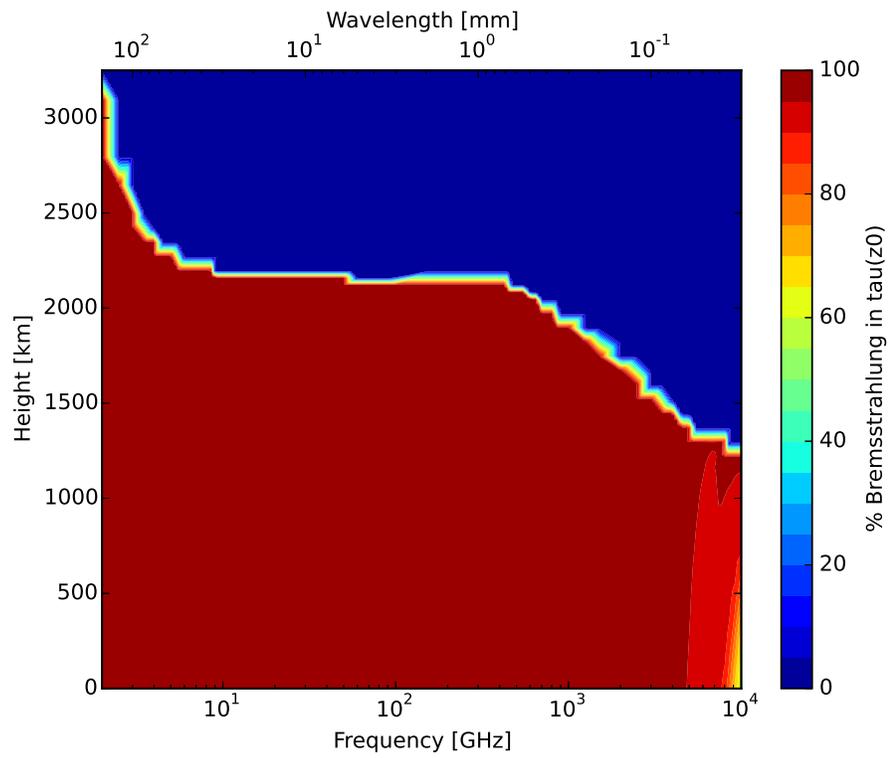}
\end{center}
\caption{Height above the limb versus frequency, in colors the percentage of contribution of Bremsstrahlung at $z_0$. For frequencies lower than 5 THz the emission is generated by  Bremsstrahlung process.}\label{h-vs-nu-vs-contribution.eps}
\end{figure}

The Figure \ref{h-vs-nu-vs-contribution.eps} shows the percentage of Bremsstrahlung contribution in the total opacity at $z_0(\nu,y)$. We found that all the radio - infrared emission at the limb is originated by Bremsstrahlung process, even at high frequencies.

\begin{figure}
\begin{center}
\includegraphics*[width=1.0\textwidth]{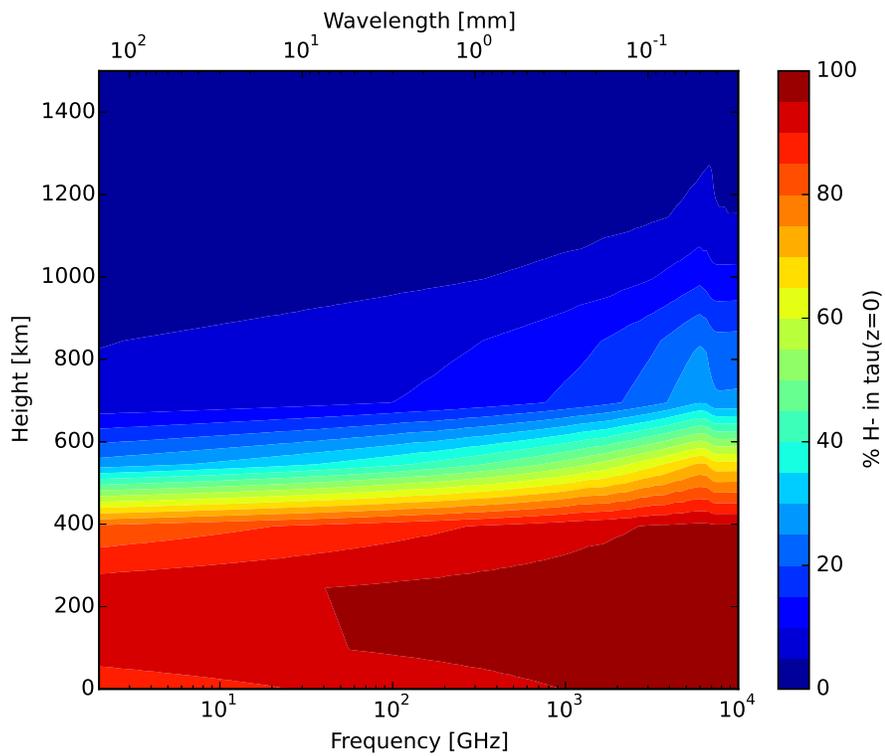}
\end{center}
\caption{Height above the limb versus frequency, in colors the percentage of H- in the total contribution of the opacity at $z=0$ km. H- contributes more than 50\%  for altitudes above the limb between $0$ and $500$ km.}\label{h-vs-nu-vs-contribution-in-z=0.eps}
\end{figure}

We analyzed the case of the opacity at $z=0$ to compare with our results for $z_0$. The Figure \ref{h-vs-nu-vs-contribution-in-z=0.eps} shows the contribution of H- in the total opacity. In this case, H- contributes 50\% or more for altitudes lower than $500$ km above the limb for all the frequencies. For high altitude the contribution of H- decrease slowly. 

\section{Conclusions}
Using the C07 model as atmospheric input conditions, we computed the optical depth and the opacity contribution (for Bremsstrahlung and H- processes) at solar limb altitudes for the millimeter - infrared spectral range. We showed that the theoretical tomography property for limb altitudes remains valid at this wavelength range. We also showed that although the Bremsstrahlung process is the main responsible of the final emission at $z_0(\nu,y)$ the H- emission plays a significant role in the emission and absorption process around depths at $z=0$. The results also showed that for altitudes higher than $500$ km above the limb, the Bremsstrahlung is the dominant emission throughout the ray path integration (over the z-axis).

We compared observations in H$\alpha$ from the spicules at the limb \citep{2014ApJ...795L..23S} and observations at sub-millimeter wavelengths \citep{1994IAUS..154..139C} against theoretical emission. We found differences of 4 Mm (for H$\alpha$ observations) and 12 Mm at 20 GHz \citep{DelaLuz2015}.

These differences are significant due the fact that Chromospheric altitudes computed with semi-empirical models are around 2.5 Mm. However, at high frequencies the observations and the models predict similar radii.

We conclude that the extended emission at low frequencies observed in \cite{1994IAUS..154..139C} could be explained with a extended Chromosphere that includes only Bremsstrahlung emission, i.e. temperatures higher than 10,000 K (Hydrogen temperature of ionization at Chromospheric conditions) with densities of around 1e11 cm${}^{-1}$ in a wider Chromosphere than the computed in the center of the solar disk.

\section{Acknowledgments}
Thanks for the program of CONACyT Fellowship (project 1045) to support this work.

\section{Citations}

\end{document}